\documentclass{PoS}

\title{Overcoming strong metastabilities with the LLR method}

\ShortTitle{Overcoming strong metastabilities with the LLR method}

\author{\speaker{Biagio Lucini}\\%
      Department of Mathematics, Swansea University, Singleton Park, Swansea SA2 8PP, UK\\
      E-mail: \email{b.lucini@swansea.ac.uk}}

\author{William Fall\\
      Department of Physics and Astronomy, University of Sheffield, Hounsfield Road,\\ Sheffield S3 7RH, UK\\	
      Department of Physics, Swansea University, Singleton Park, Swansea SA2 8PP, UK\\
      E-mail: \email{william.fall@sheffield.ac.uk}}

\author{Kurt Langfeld\\
        Department of Mathematical Sciences, University of Liverpool, Liverpool L69 7ZL, UK\\
        Centre for Mathematical Sciences, Plymouth University,
        Plymouth, PL4 8AA, UK\\
        E-mail: \email{Kurt.Langfeld@liverpool.ac.uk}}

\abstract{In previous work, it has been shown that the recently
  proposed LLR method is very efficient at overcoming strong
  metastabilities that arise near first-order phase transition
  points. Here we present a systematic study of the performance of the
  algorithm near (pseudo-)critical points for $q$-state Potts models with $q$
  as large as 20, in two and three dimensions. In particular, we shall
  focus our study on the ergodicity of the replica exchange step and
  the underlying physical mechanism. When compared with both
  analytical and numerical results present in the literature, our
  determinations of thermodynamic observables (including the
  order-disorder interface tension at criticality) show an impressive
  degree of relative accuracy (up to $2.5 \times 10^{-6}$), which confirms the
  reliability and the efficiency of the proposed approach.}

\FullConference{34th annual International Symposium on Lattice Field Theory\\
		 24-30 July 2016\\
		 University of Southampton, UK}

\begin{document}

\section{Introduction: the LLR algorithm and ergodicity through replica exchange}
The LLR algorithm was originally introduced in~\cite{Langfeld:2012ah} as an extension of the ideas behind the
Wang-Landau sampling~\cite{Wang:2000fzi} allowing us to efficiently extract the
density of states in systems with a continuous energy spectrum. Following
further developments, this algorithm has shown a range of
applicability that encompasses a rich variety
of problems in lattice gauge theories and statistical mechanics,
including exponentially growing tunneling times at first order phase
transitions~\cite{Langfeld:2015fua} and the sign
problem~\cite{Langfeld:2014nta}.  

The LLR algorithm determines numerically the density of states of
a system using a non-Markovian sampling. A key aspect of the approach
is the sequencing of the whole energy spectrum in intervals of
amplitude $\delta$, assumed to be small enough for the logarithm of the
density of states to be accurately approximated by its first order Taylor
expansion. The latter is then determined with a recursive relation
including steps of standard Monte Carlo
sampling, with the constraint that the system never leaves the given
energy interval. A review of the method, which we use in the formulation
given in~\cite{Langfeld:2015fua}, has been provided at this
conference~\cite{Langfeld:Lat16}. Here we focus our attention on the
energy restricted Monte Carlo. 
One of the potential problems with restricting the energy interval
consists in a possible loss of ergodicity, which would invalidate the
numerical results obtained with the algorithm. While one
might assume that the restricted energy Monte Carlo explores all the
configurations associated with the given energy range, that is not
necessarily the case. In fact, there are scenarios in which the
configuration space can become disconnected if one restricts the sampling
in the energy (e.g. when the system has some non-trivial topological sectors). 

\begin{figure}
\begin{center}
\includegraphics[width=.65\textwidth]{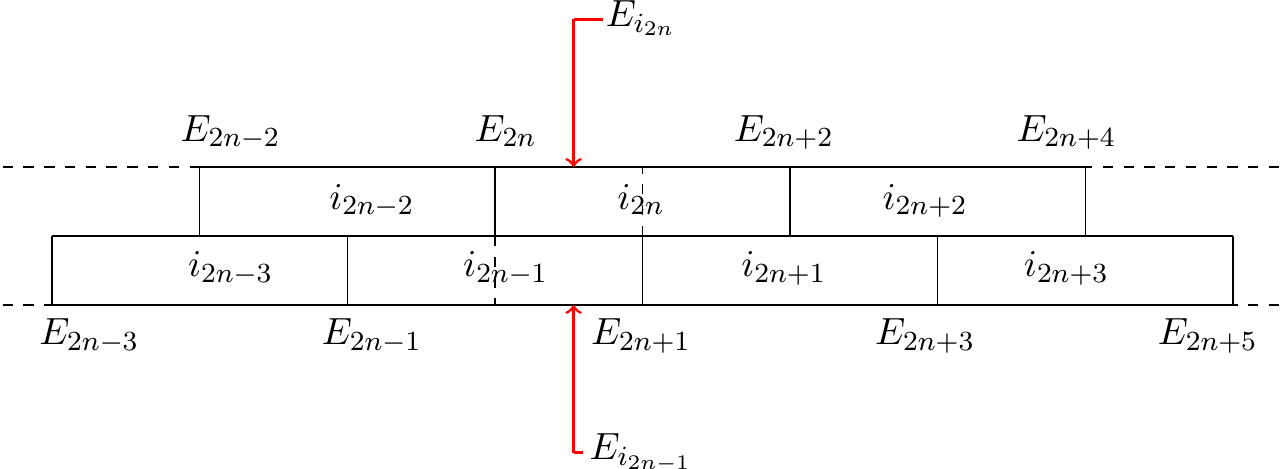}
\end{center}
\caption{The two sequences of overlapping intervals covering the
  whole energy range. The case of a coincidence of two energies
  $E_{i_{2n}}$ and $E_{i_{2n}-1}$ in an overlapping region is also
  sketched.\label{fig:replica_sketch}} 
\end{figure}

In order to overcome this problem,
in~\cite{Langfeld:2015fua} a replica exchange step has been
proposed\footnote{An alternative method, in which the sharp
  restriction in a given interval is replaced with the convolution
  with a Gaussian, has been discussed in~\cite{Pellegrini:Lat16}.}. The proposal consists in
having a second sequence of energy intervals displaced from the
original one by half an interval (Fig.~\ref{fig:replica_sketch}). After a certain number of
restricted Monte Carlo steps, when in two overlapping intervals (which
we call $i_{2n-1}$ and $i_{2n}$) both configurations are in the
overlapping region, one proposes a swap of configurations with a
Metropolis probability 
\begin{eqnarray}
P_{\mathrm{swap}} = \mathrm{min}\left(1,e^{\left(a^{(m)}_{i_{2n}} -
      a^{(m)}_{i_{2n-1}}\right) 
  \left(E_{i_{2n}} - E_{i_{2n-1}}\right)}\right)  \ . 
\end{eqnarray}
In this expression, $a^{(m)}_{i_{2n}}$ (respectively, $a^{(m)}_{i_{2n-1}}$) is the current iterative estimate of the first
Taylor coefficient in the interval $i_{2n}$ ($i_{2n-1}$) and
$E_{i_{2n}}$  ($E_{i_{2n-1}}$) the energy of the configuration in $i_{2n}$ ($i_{2n-1}$). 
Via subsequent swaps, the system can travel through the whole
energy range, hence solving any potential ergodicity issue. 
A test that would enable us to verify whether this idea works in
practice would be to establish that at convergence the system moves
across intervals following a random walk, as it happens in the
Wang-Landau algorithm. 

In this work, we study numerically the replica exchange idea for the
first time, using as a prototype the $q$-state Potts model
(see~\cite{Guagnelli:2012dk} for a formulation of the LLR method for
spin systems) in two and three dimensions and for $q$ as large as
20. 

\section{Application to the Potts model}
\begin{figure}[tb]
\begin{tabular}{cc}
\includegraphics[width=.49\textwidth]{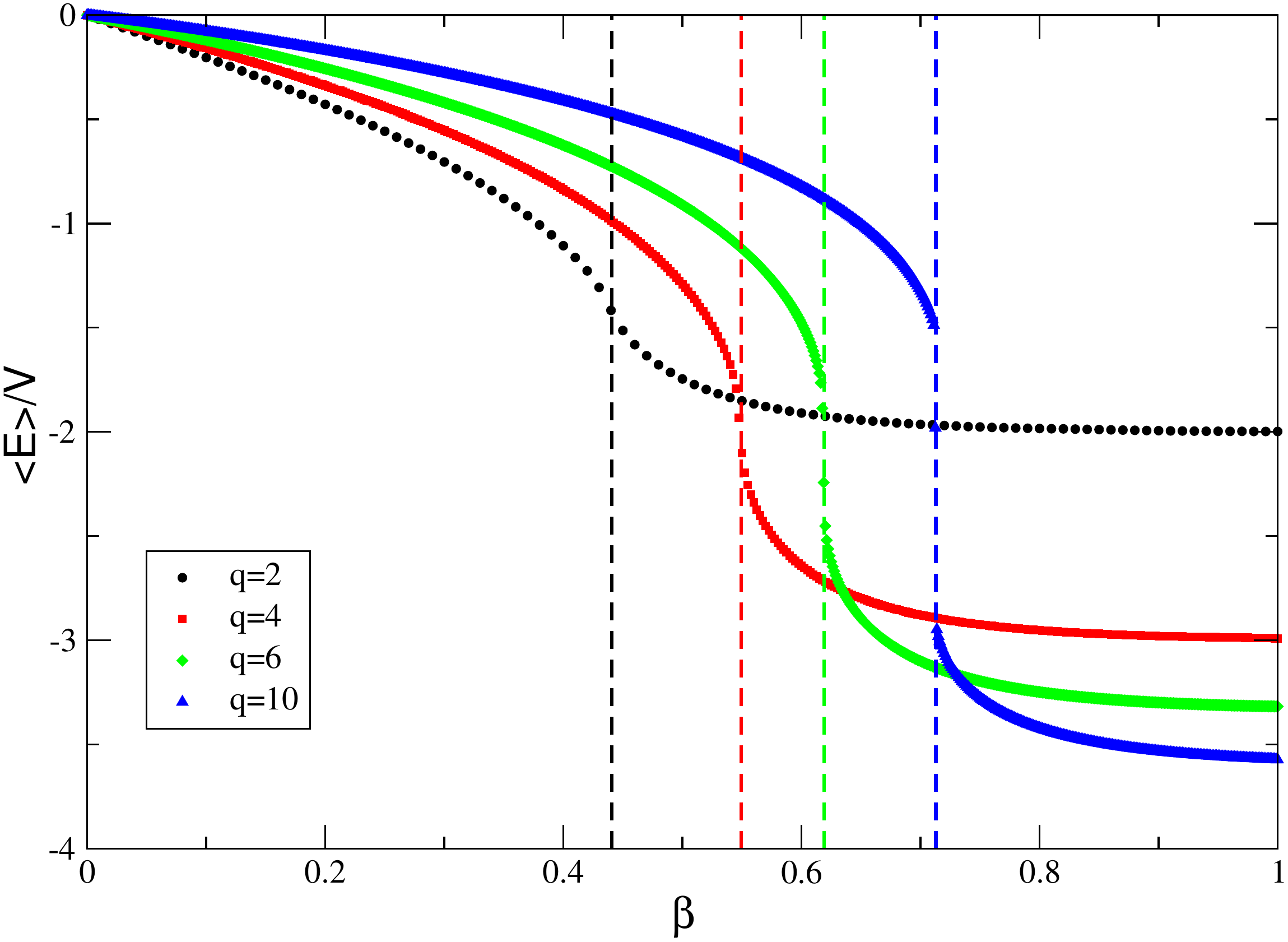} &
\includegraphics[width=.49\textwidth]{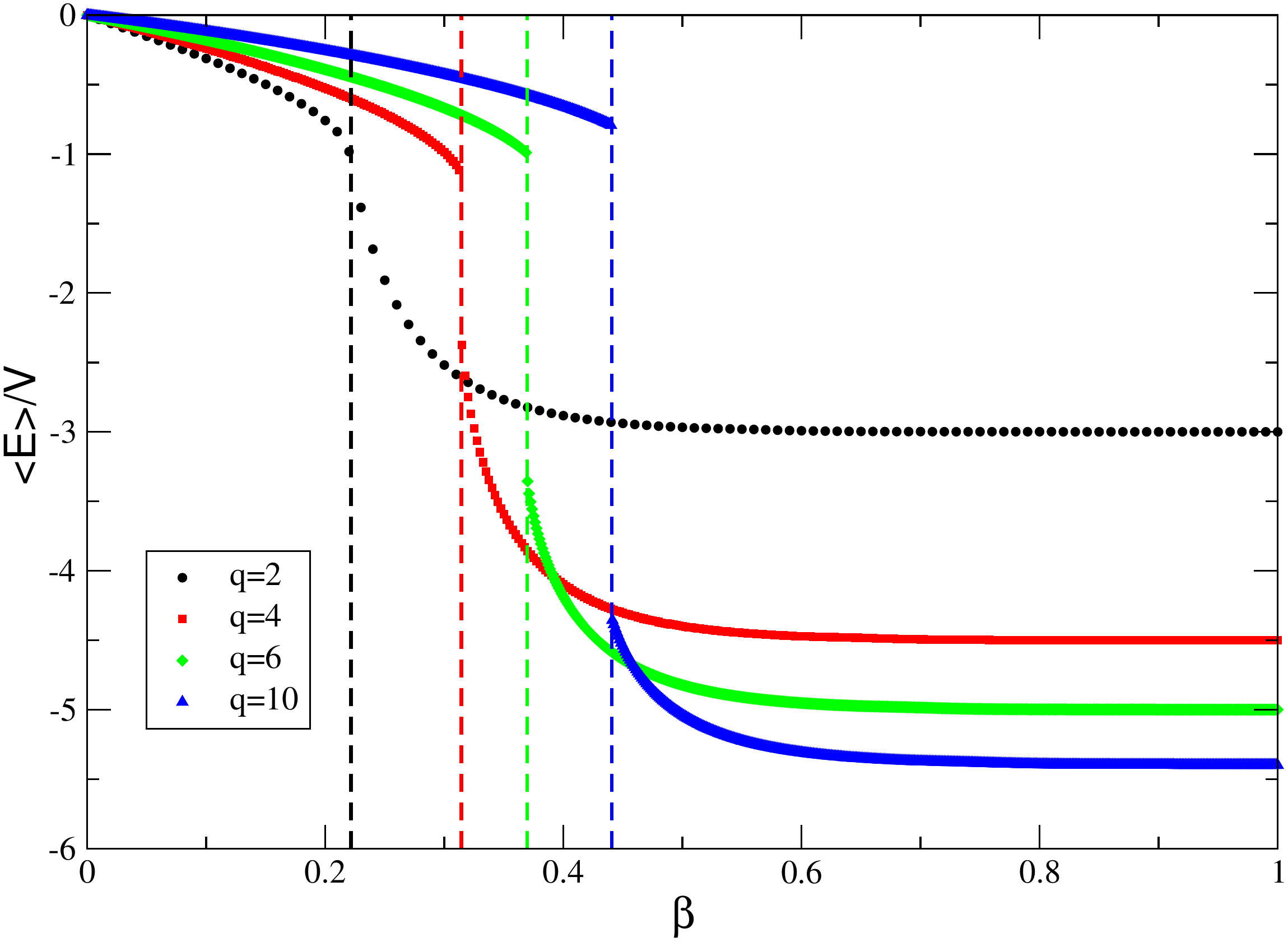}
\end{tabular}
\caption{The average energy per unit site as a function of $\beta$ for
  $D=2$, $L=64$ (left) and $D=3$, $L=16$
  (right), at the shown $q$ values. The vertical lines
  indicate the position of $\beta_c$. \label{fig:evsbeta}}  
\end{figure}
The Hamiltonian of the $q$-state Potts model in $D$ dimensions is given by
\begin{equation}
H = 2 \beta \sum_{\langle ij \rangle} \left( \frac{1}{q} -
  \delta_{\sigma_i, \sigma_j} \right) \ , 
\end{equation}
with the spin variables $\sigma_i$ taking  the values $0, \dots, q-1$ and the sum ranging
over all lattice links. The lattice is symmetric, with linear size $L$
and volume $V=L^D$. 

For $D>1$, as a function of the coupling $\beta$,  the system has a 
$\mathbb{Z}_q$-driven phase transition. In $D=2$ the $q$-state Potts
model (which for $q=2$ is equivalent to the Ising model) has a first
order phase transition for 
$q>4$, while in $D=3$ the transition is first order for $q >2$. In both
cases, the strength of the transition grows significantly with $q$. In
$D=2$ the $q=10$ case is generally taken as a benchmark of
efficiency of algorithms~\cite{Wang:2000fzi}, while the highest value of $q$
simulated we are aware of is $q=30$~\cite{Binder:2012}. For $D=3$
accurate Multicanonical Monte Carlo results have been presented
in~\cite{Bazavov:2008qg} up to $q=10$. Both known analytic results
(for $D=2$) and precise numerical calculations already reported in the
literature provide a solid reference base against which novel numerical
approaches like ours can be assessed. 

As a first check of our method, we verify that the energy as a
function of $\beta$ has the expected behaviour, i.e. it goes to zero
in the limit $\beta \to 0$ and to $4(1 - q)/q$ in the limit $\beta \to
\infty$, with clear indication of the phase transition arising at
$\beta_c$. This is shown in Fig.~\ref{fig:evsbeta}, in which we plot
the energy per unit site as a function of $\beta$ for $q =
2,4,6,10$. In $D=2$ we have used square lattices of linear size $L=64$,
while in $D=3$ we have simulated cubic lattices with $L=16$. The
critical values of $\beta$ used for this test, also reported in the
figure, are the exact infinite size values in two dimensions,  
\begin{equation}
\label{eq:betac}
\beta_c = \frac{1}{2} \log \left( 1 + \sqrt{q} \right) \ ,
\end{equation}
and the numerical extrapolations to $L = \infty$  reported
in~\cite{Bazavov:2008qg} for three dimensions. The 
figure supports the correctness of our simulation approach. In
particular, the would-be infinite volume
singularity at $\beta_c$ is well visible, with clear indication of
a jump in the thermodynamic limit at the largest values of $q$.

\begin{figure}[tb]
\begin{tabular}{cc}
\includegraphics[width=.45\textwidth]{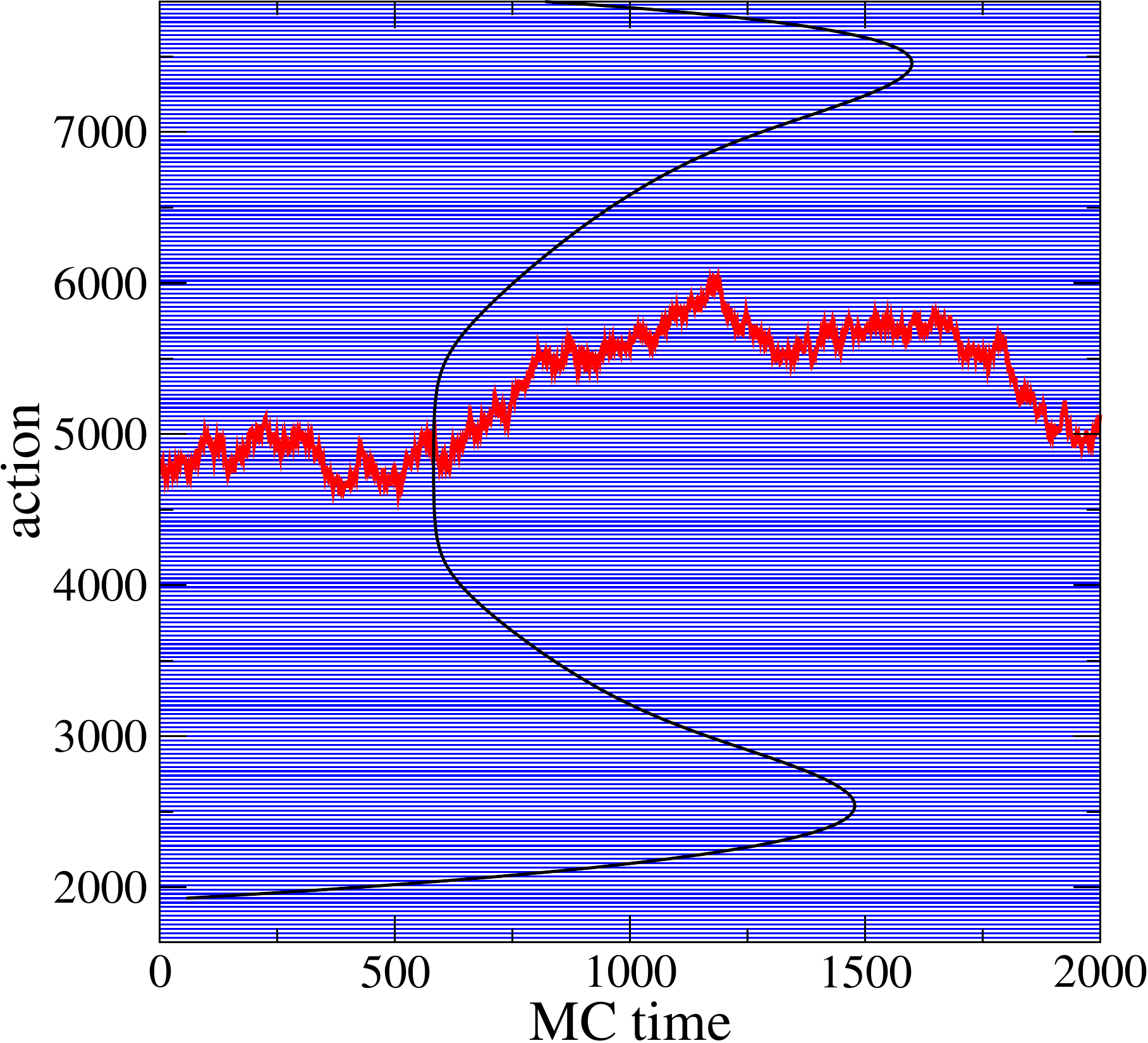} &
\includegraphics[width=.45\textwidth]{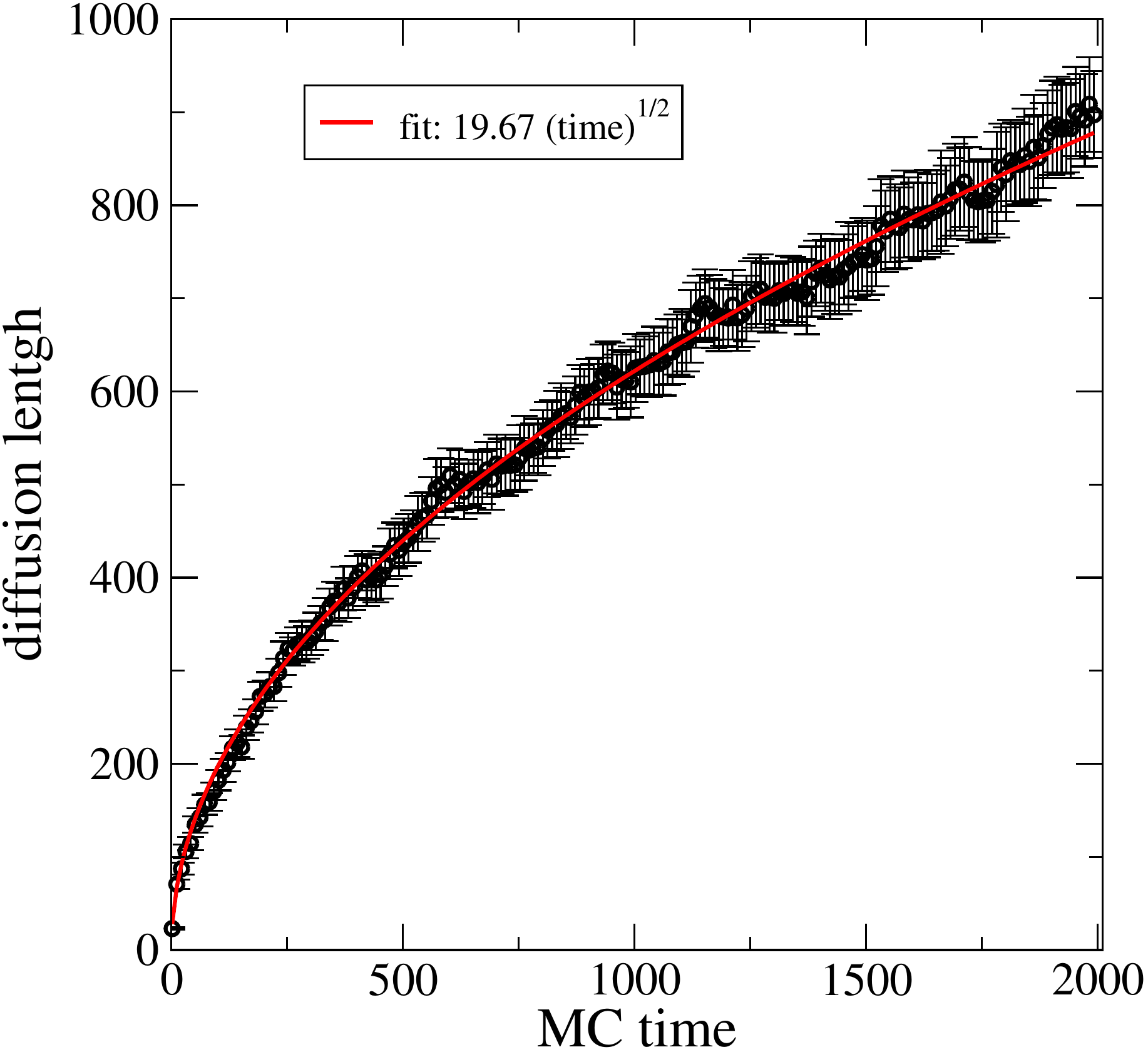}
\end{tabular}
\caption{{\em Left}: A Monte Carlo history of the action $A =
  \sum_{\langle ij     \rangle} \delta_{\sigma_i,     \sigma_j} $ of the system evolved from
  a configuration in the displayed central energy interval with the replica
  exchange step enabled. The dark and light blue bands visualise the
  energy   intervals. {\em Right}: Quadratic fit of the diffusion
  length as a function of the Monte Carlo time. \label{fig:replica}} 
\end{figure}

Next, we analyse more directly the dynamics provided by the replica
exchange. As a first investigation, we follow the history of a
configuration originally having energy in an interval that is
in between the two peak values at criticality. In this
region of energies, an interface separating the two phases
(the ordered and the disordered phase) develops near the critical
temperature\footnote{Because of the presence of both phases, this region is
  called the {\em mixed phase} region.}. This study is performed for $q=20$ and $D=2$. As the 
replica exchange step is turned on, the system is found to be travelling in energy
space in a fashion that is clearly reminiscent of a random walk
(Fig.~\ref{fig:replica}). We then fit the diffusion mean length $E_d =
\langle (E - E_0)^2 \rangle^{1/2}$ using the random 
walk law 
\begin{equation}
E_d = C_d t_{MC}^{1/2} \ , 
\end{equation}
where $t_{MC}$ is the Monte Carlo time and $C_d$ the diffusion
coefficient. In this particular case, we find $C_d\simeq 19.67$. The
quality of the fit is displayed in Fig.~\ref{fig:replica}, right. $C_d$
seems to be independent of the starting energy: for an energy that
corresponds to an equilibrium value in the ordered phase we find $C_d \simeq
18.73$, while for an energy corresponding to an equilibrium value in
the disordered phase we find $C_d \simeq 19.80$. This suggests that our
algorithm produces a random walk over all energies, hence avoiding
any trapping that can result from the original LLR prescription. 

\begin{figure}[tb]
\begin{tabular}{cc}
\includegraphics[width=.49\textwidth]{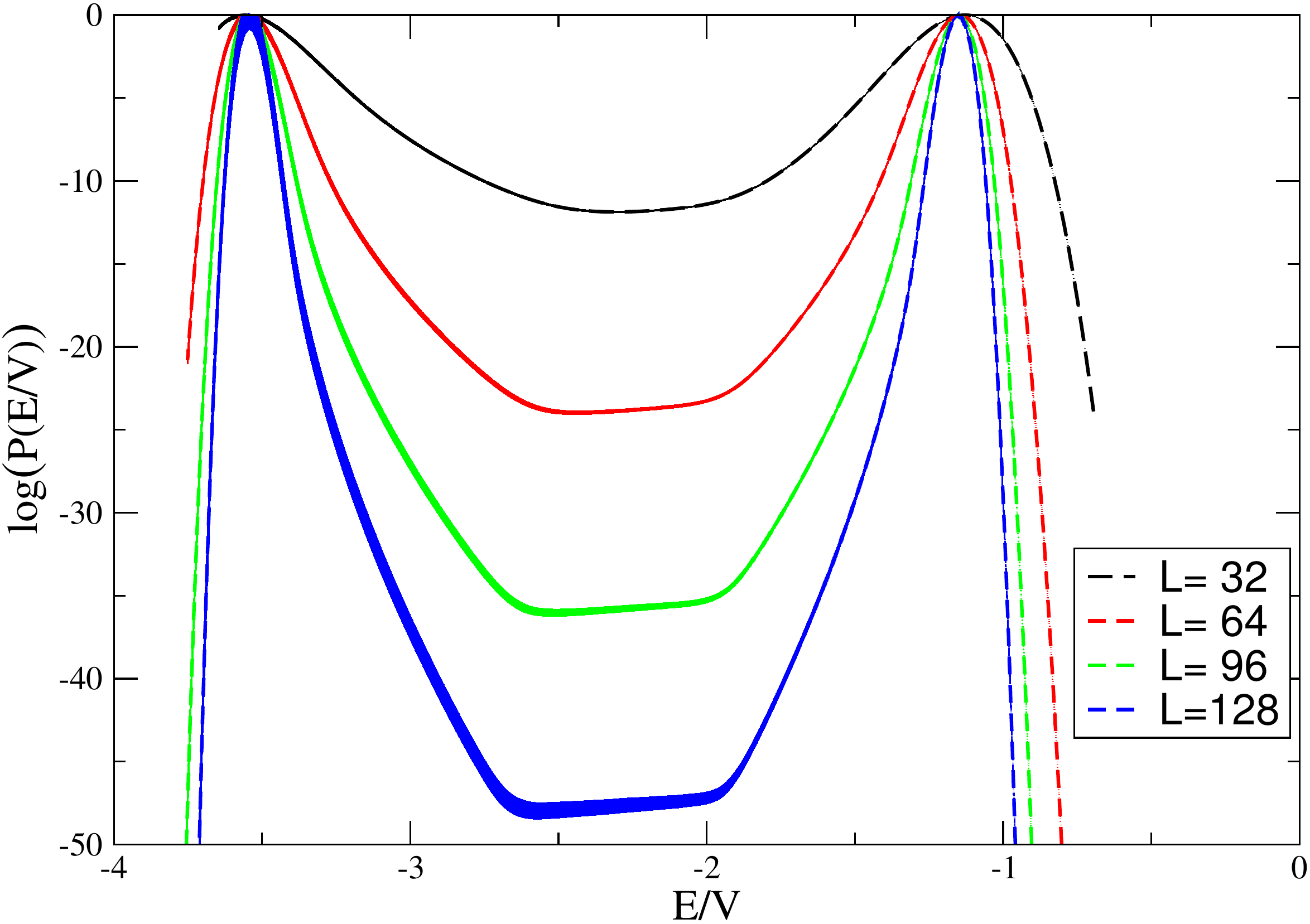} &
\includegraphics[width=.49\textwidth]{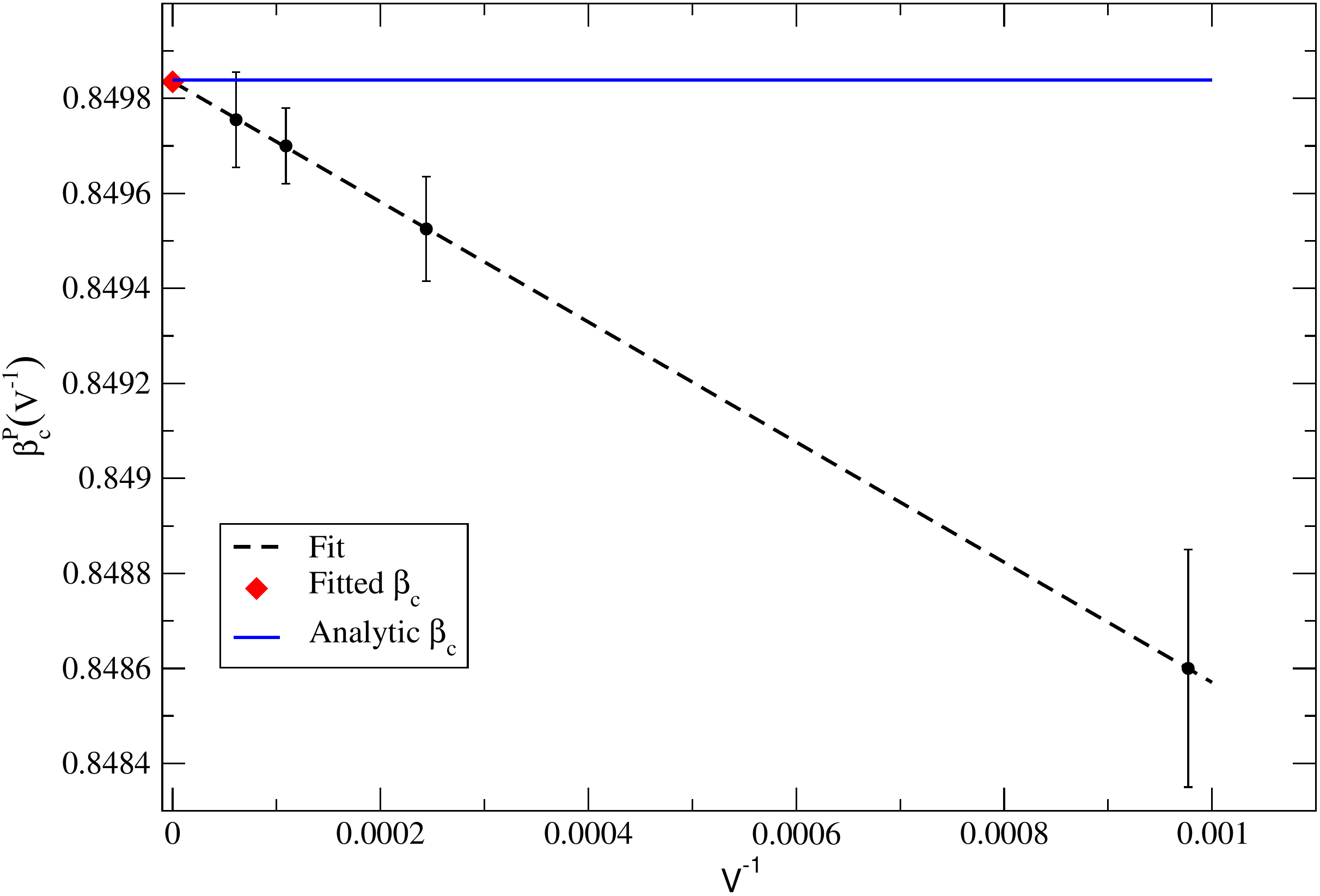}
\end{tabular}
\caption{ {\em Left}: The logarithm of the probability distribution of the
  energy per unit site as a function of the latter quantity at
  $\beta_c^P(L)$. {\em Right}: Infinite volume extrapolation of
  $\beta_c^P(V^{-1}) \equiv \beta_c^{P}(L=V^{1/2})$. \label{fig:logrho}}
\end{figure}

We then move to a quantitative analysis of the phase transition. Also for
this study, we consider the Potts model with $q=20$. Due to the severe
metastabilities in the critical region, this investigation is expected
to provide a compelling test of the efficiency of the algorithm at
first order phase transition points. We start our numerical study from the
identification of the critical value of the coupling $\beta_c$. At
finite size, various definitions of $\beta_c(L)$ are possible, which
do not necessarily identify the same value at fixed $L$, but 
all converge  to a unique $\beta_c$ when $L \to \infty$. Here, we choose to
identify $\beta_c(L)$ with $\beta_c^P(L)$, the value of $\beta$ at which the
probability distribution of the energy per unit site $P(E/V)$ has two peaks of
equal height\footnote{See~\cite{Langfeld:Lat16} for an alternative
  definition.}. A plot in logarithmic scale of $P(E/V)$ is reported 
in Fig.~\ref{fig:logrho} (left). We note the presence of a flat valley
between the two maxima, whose depth increases as $L$ is
increased. The latter is the feature that ultimately leads to the
failure of local update algorithms at first order phase transitions,
since in order to travel through the valley one would need a
simulation time that grows exponentially with $L^{D-1}$. The visual
indication from the plot is that our algorithm does allow us
to sample efficiently the metastable vacua and the mixed phase
configurations that characterise the valley.

In order to have a more stringent quantitative test, we extract the
critical values $\beta_c(L)$ using a bootstrap analysis on
20 different simulations. The latter have been performed feeding
independent Markov chains to the LLR algorithm.

The system having a first order phase transition, finite size scaling predicts
that
\begin{equation}
\beta_c^P(L) = \beta_c + \frac{A}{L^D} \ , 
\end{equation}
with $A$ a constant. Our determinations of $\beta_c(L)$ are displayed in the right
pane of Fig.~\ref{fig:logrho} together with the expected linear
dependency in $V^{-1} = L^{-D}$. Beside an obvious overestimate of the errors (whose
understanding is a work in progress), the excellent quality of the fit
is evident. The extracted numerical value is $\beta_c^{fit} =
0.8498350(21)$, which is perfectly compatible with the $q=20$ value
coming from the exact expression, Eq.~(\ref{eq:betac} ). More in
details,
\begin{equation}
\frac{\beta_c^{fit} -  \beta_c^{exact}}{\beta_c^{exact}} = 1.7(2.5) \times 10^{-6} \ ,
\end{equation}
i.e. the compatibility is verified up to a statistical error that
is 2.5 parts per million. Not only does this analysis demonstrate the
efficiency of the algorithm, but it also shows the high level of
accuracy that it allows us to obtain. 

\begin{figure}[tb]
\begin{center}
\includegraphics[width=.55\textwidth]{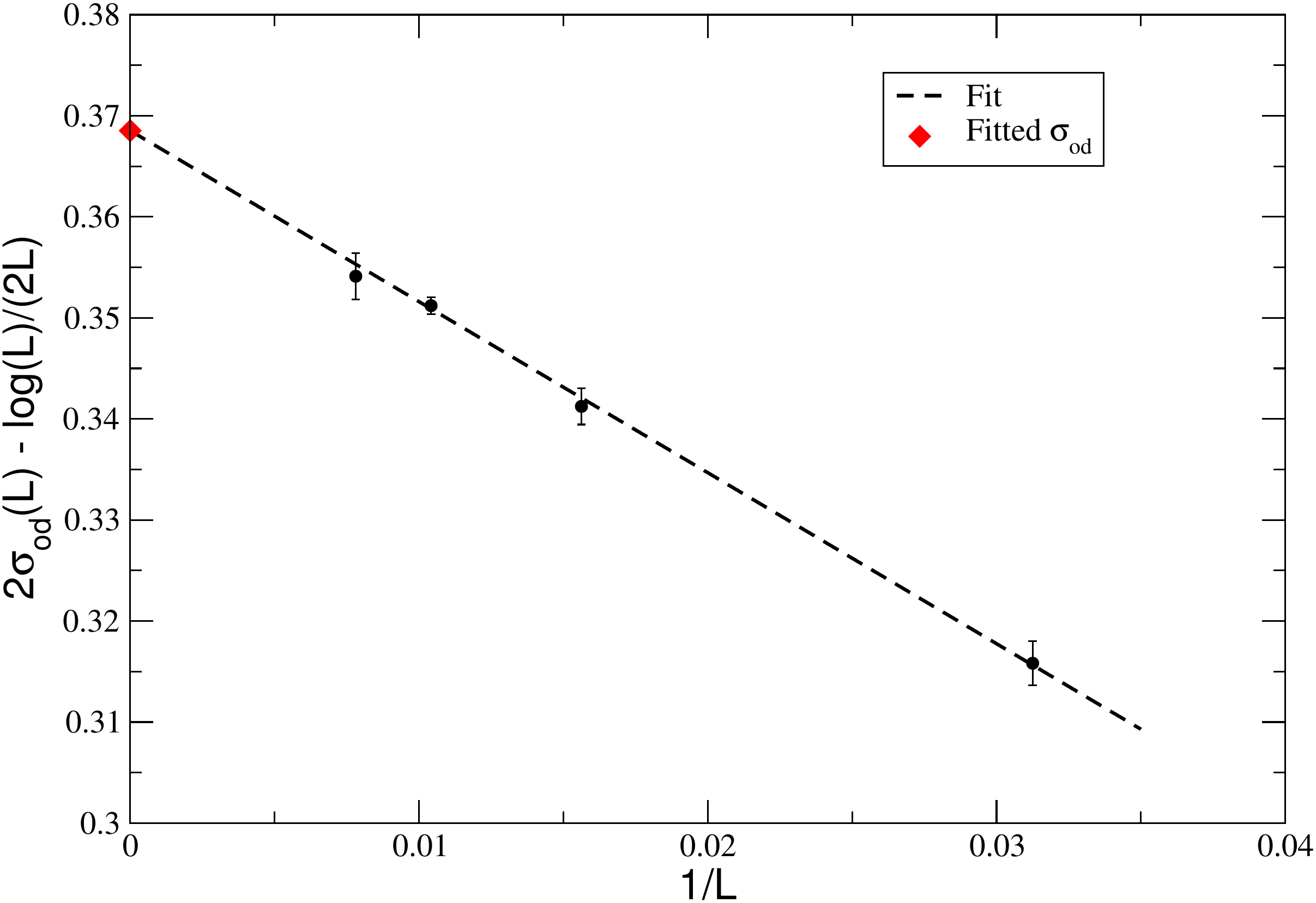}
\end{center}
\caption{The finite-$L$ interface tension $\sigma_{od}(L)$
  as a function of the size for $q=20$ and $D=2$. The infinite $L$
  extrapolation is also provided. \label{fig:interface}} 
\end{figure}

The tension of the interface between ordered and disordered
phase is an observable that plays an important role at first order
phase transitions (see e.g.~\cite{Neuhaus:2002}). This quantity can be extracted
from the relative height between the two equal peaks and the minimum in the flat
region between them at $\beta_c^P(L)$. In particular, if we define the interface
tension $\sigma_{od}$ from 
\begin{equation}
2 \sigma_{od}(L) = - \frac{1}{L} \log P_{min, valley} \ ,
\end{equation}
where the peak probability has been normalised to one, the latter can
be extrapolated to infinite $L$ using the ansatz
\begin{equation}
\label{eq:intextr}
2 \sigma_{od} (L) - \frac{\log L}{2 L} = 2 \sigma_{od} +
\frac{c_{\sigma}}{L}  \ ,
\end{equation}
where $\frac{\log L}{2 L} $ is an entropic term due to the fact that
the system is translationally invariant along the direction of the
interface. Our data are displayed in Fig.~\ref{fig:interface}, where
we show also the best fit according to~(\ref{eq:intextr}). For the infinite
volume value, we obtain $2 \sigma_{od} =
0.36853(88)$. This can be compared to the strong coupling analytic
result of Borgs and Janke~\cite{Borgs:1992}, which, for $q=20$, gives
$2 \sigma_{od} = 0.3709881649\dots$. The relative 
discrepancy of the two results is of the order of $7 \times
10^{-3}$. While there is no reason to expect perfect agreement, since
the analytic result is not exact (although is known to provide a very
good approximation to the exact value) and the numerical result is
still preliminary, it is reassuring for our purposes that the numbers
are close at the sub-percent level. Further analysis will be performed
to check the robustness of our determination.

\section{Conclusions}
In this contribution, we have shown that the LLR method provides an
efficient numerical tool for investigating strong metastabilities at
first order phase transitions even in the case of the Potts model, for
which the energy spectrum is discrete. In particular, the replica exchange
method plays a crucial role in extending the random walk from the
single interval in which the simulation is requested to be restricted
by the LLR original prescription to the whole energy range. The
provided verification of
the random walk dynamic of the Monte Carlo time series supports the
conjecture that the algorithm scales as $V^2$ also in the case of a
first order phase transition. This has to be contrasted with Markov
Chain Monte Carlo methods, for which the scaling is known to be
exponential in $L^{d-1}$. It is worth remarking that the results
presented here have been obtained with modest computational times
(three weeks of runs on around 200 cores of state of the art processors).
We regard this study as preliminary. A detailed study of
some aspects only hinted in this work (among which is
the scaling of the algorithm with the system size) is the focus of our
current investigation and will be reported elsewhere.

\acknowledgments{
We thank E. Bennett, N. Garron, R. Pellegrini and A. Rago for discussions. The
work of BL is partially supported by the STFC Consolidated Grant
ST/L000369/1. KL is supported in part by the Leverhulme Trust (grant
RPG-2014-118) and STFC (grant ST/L000350/1). Numerical computations
were executed  at the HPCC in Plymouth and on the HPC Wales systems,
supported by the ERDF through the WEFO (part of the Welsh Government).  
}

\end{document}